\documentstyle[12pt]{article}

\title{
Exact results on linear response of cyclic molecular aggregates}

\author{
A.~V.~Ilinskaia$^{1}$\thanks{E-mail: ILINSKI@PHIM.NIIF.SPB.SU},
K.~N.~Ilinski$^{2,3}$\thanks{E-mail: KNI@TH.PH.BHAM.AC.UK},
G.~V.~Kalinin$^{1}$\thanks{E-mail: KALININ@PHIM.NIIF.SPB.SU},
\\
V.~V.~Melezhik$^{1}$\thanks{E-mail: MELEZHIK@PHIM.NIIF.SPB.SU},
A.~S.~Stepanenko$^{4,3}$\thanks{E-mail: STEPANENKO@PHIM.NIIF.SPB.SU}
\\
{\small\it $^{1}$ Institute of
Physics, Physics Department of St-Petersburg University,}
\\
{\small\it St-Petersburg, 198904, Russian Federation.}
\\
{\small\it $^{2}$ School of Physics and Space Research, University of
Birmingham,}
\\
{\small\it Birmingham B15 2TT, United Kingdom.}
\\
{\small\it $^{3}$ Institute of Spectroscopy, Russian Academy of
Sciences,}
\\
{\small\it Troitsk, Moscow region, 142092, Russian Federation.}
\\
{\small\it $^{4}$ St-Petersburg Nuclear Physics Institute,
Gatchina,}
\\
{\small\it St-Petersburg, 188350, Russian Federation.}}
\date{ }

\begin{document}

\maketitle

\setcounter{page}{0}
\thispagestyle{empty}
\vskip -12cm
\rightline{preprint TPBU-8-95}
\rightline{cond-mat/9509040}
\vskip 12cm

\begin{abstract}
Basing on the picture of Frenkel excitons in molecular crystals
described by the XY-model we
consider the linear response of linear cyclic aggregates at finite
temperature.  The exact results for characteristics of the
response are obtained.  In particular, we calculate time-dependent
two-point correlation functions at finite temperature for the
cyclic 1-D XY-model.
\end{abstract}
\newpage

\section{Introduction}

Last five years a number of papers were devoted to the consideration
of the so-called J-aggregates, molecular linear aggregates with
redshifted unusually sharp absorption bands (for the recent review
see \cite{Knoester}).  Such aggregates give the infrequent
possibility to describe a nontrivial physical situation by some sort
of 1-dimensional quantum Hamiltonians which could be treated exactly.
To be more precise, the optical excitations of the system (excitons)
after the second quantization procedure could be described by the
Hamiltonian of spin chains (we will see in section 2 that this is
the anysotropic XY-chain).  Moreover, theoretical investigations of
J-aggregates essentially use the information about the structure of
the exact eigenvalues and eigenfunctions of the Hamiltonian which are
well-known due to the paper by Lieb, Mattis and Schults \cite{LMS}
since 1961.

The purpose of this paper is to improve the results on the linear
response \cite{Note1} of the J-aggregates in the following aspects:
\begin{enumerate}
\item
As we will remind in the next section, there is no freedom to
restrict the consideration from the XY-model to the XX-model, because
the relative anysotropy parameter is the order of unit.  This fact
causes a number of complications which are not only technical ones.
Indeed, there is no more good quantum number like the number of
particles in the chain (what is obvious from physical features of
the problem) and one has to deal with a full Fock space of
particles in question.
\item
The same problem (i.e. the calculations in a full Fock space of the
theory) appears even in the  XX-model when the effects of
nonzero temperature are considered and the correlation functions are
some traces on a full Fock space.
\end{enumerate}
Both of these problems are treated in the paper.

The next question of interest is a choice of the boundary conditions
for finite linear aggregates.  It is well understood that the choice
of the general closed boundary conditions leads to the new nontrivial
problem.  Indeed, the exchange statistics for the excitons (which are
statistically Paulions, i.e. fermions on site and bosons on different
sites) is irrelevant for the open boundary conditions because no real
permutation is allowed.  The situation changes if the linear
aggregate is a circle.  It leads to a nontrivial loop which gives the
possibility to permute particles via the 'glued' boundary.  Then the
effects of the exchange statistics exhibit themselves \cite{IKK}.
These effects are negligible in the continuous limit and were not
investigated when the main interest was concentrated on the
thermodynamics of the chains.  A large number of the classical and
recent papers concern the subjects \cite{N,BM,Iz1,Iz2}.  But the
effect of the closure is relevant and important for  finite
systems like J-aggregates.  Moreover, they allow to trace the
crossover from small to macroscopic chains \cite{Knoester}.  Namely
the cyclic finite aggregates will be considered in this paper.

We have to note that the exact treatment of excitons in the
aggregates on the base of the cyclic XX-model were considered in
\cite{Jap}.  In Section~3.4 we will criticize the results of
Ref.\cite{Jap}.

The paper is organized as follows.  In the next section we give a
brief description  of  excitons in molecular chains which are
basic for understanding of the subject.  While opticians began to
deal with  excitons long time ago, mathematical
physicists (who are mainly concentrated on exact solutions) are not
so familiar with the field.  The section can be considered as an
introduction in the excitonic physics.  It also contains the formula
for the linear response function in terms of correlators of the
system in the convenient for the following form.  Section~3 contains
a main new result of the paper.  It is devoted to the exact
calculation of two-point correlation functions for the cyclic
XY-model in a closed form.    The
conclusion completes the paper with a discussion of the resulting
formulae for  characteristics of the linear response for the
cyclic J-aggregates.  Appendix~A contains the evaluation of the main
integrals of Section~3.

\section{Linear response of excitonic system}

In the present work we will calculate the optical response of a gas
of molecular aggregates.  We are interested in the average dipole
momentum induced by an external time dependent electromagnetic field
in the linear approximation with respect to the field amplitude .

The most general linear relation between the dipole momentum and the
field has the form
$$
{p}_\alpha(t) =
\int\limits_{0}^{\infty}
\tilde{\sigma}_{\alpha\beta}(\tau)
E_\beta (t-\tau) \mbox{d}\tau
$$
where ${p}_\alpha$ is the dipole momentum per unit volume,
$E_\beta$ denotes the external field,
$\tilde{\sigma}_{\alpha\beta}(\tau)$ is the linear response function
and $\alpha$, $\beta$ are tensor indices.

We suppose that the aggregates are independent, i.e.  we will neglect
the interaction of the aggregates.  In this case the dipole momentum
per unit volume is just a product of the aggregate concentration
$\nu$ and the dipole momentum of a single aggregate. Hence the
response $\tilde{\sigma}_{\alpha\beta}$ is also proportional to the
concentration $\nu$:
$$
\tilde{\sigma}_{\alpha\beta}(\tau) =
\nu\sigma_{\alpha\beta}(\tau)
$$
where $\sigma_{\alpha\beta}(\tau)$ is the response function of a
single aggregate.

As the model for a molecular aggregate we consider the 1-D closed
M-sites molecular chain (that can be regarded for example as a model
for the benzene vapor under a low pressure).  Let us derive the
excitonic Hamiltonian of the aggregate following to the original
paper \cite{A0}. We start with the Hamiltonian of the molecular
aggregate in the form:
\equation
\label{agregate}
{\bf H} =
\sum\limits_{n=1}^{M} H_n +
\frac12\mathop{{\sum}'}\limits_{n,m} V_{n m}
\endequation
Here $H_n$ is the Hamiltonian of a molecule on the n-th site,
$V_{nm}$ is the operator the of interaction between n-th and m-th
sites, $\sum '$ means the absence of the term with $n=m$.

As far as the molecules are supposed to be electrically neutral, the
first approximation for the interaction $V_{nm}$ is dipolar one.

Let us denote  the complete set of eigenfunctions of the free
molecular Hamiltonian $H_n$ by $\{\phi_n^f\}$  and the corresponding
occupation number by $N_{nf}$ .  After the second quantisation
procedure the  state of the aggregate becomes a function of
occupation numbers
$$
\left|
\dots
N_{nf}
\dots
\right.
\rangle
$$
and Hamiltonian (\ref{agregate}) can be rewritten by the
following way:
\equation
\label{second}
{\bf H} =
\sum\limits_{n,f} \epsilon_f b_{nf}^+ b_{nf} +
\frac12 \mathop{\mathop{{\sum}'}\limits_{n,m}}\limits_{f,f',g,g'}
        b_{nf'}^+b_{mg'}^+ b_{mg} b_{nf}
\langle
f'g'
\left|
V_{nm}
\right|
fg
\rangle
\endequation

The creation (annihilation) operators $b_{nf}^+$ ($b_{nf}$) of the
$f$-state on the n-th site satisfies the standard commutation
relations:
$$
\{ b_{nf}, b_{nf}^+\} = 1
$$

$$
b_{nf} b_{nf} = b_{nf}^+ b_{nf}^+ = 0
$$
with the constraint
(each molecule can be in a single eigenstate only),
$$
\sum_{f} b_{nf}^+ b_{nf} = 1 \ .
$$
$\epsilon_f$ is the energy of the $f$-th excited state.

Further we shall assume that the lowest excited molecular state plays
the main role in the absorption and radiation processes and therefore
the index $f$ runs only over two values 0 (ground state) and $f$ (the
first excited state).  This allows us to present the Hamiltonian in
the form:
\equation
\label{sum}
{\bf H} = {\bf H}^1+{\bf H}^2+{\bf H}^3+{\bf H}^4
\endequation
where the first term describes the energy of nonexcited molecules,

\equation
\label{H1}
{\bf H}^1 = M\epsilon_0 +
          \frac12{\sum\limits_{n,m}}'
\langle 00 \left| V_{nm} \right| 00 \rangle,
\endequation
the second term is the energy of excitation of a single molecule in
the aggregate:
$$
\array{l}
{\bf H}^2 =
\sum\limits_n
\left(
\Delta\epsilon_f + {\cal D}_f
\right)
\hat{N}_{nf}
\\
\Delta\epsilon_f = \epsilon_f - \epsilon_0
\\
{\cal D}_f =
{\sum\limits_{m}}'
\left\{
\langle 0f \left| V_{nm} \right| 0f \rangle-
\langle 00 \left| V_{nm} \right| 00 \rangle
\right\}
\endarray
$$
and the rest terms have the form:
$$
\array{l}
{\bf H}^3 =
{\mathop{\sum{}'}\limits_{n,m}}
M_{nm}^f
b_{n0}^+b_{mf}^+b_{m0}b_{nf}
\\
{\bf H}^4 =
\frac12{\mathop{\sum{}'}\limits_{n,m}}
M_{nm}^f
\left\{
b_{n0}^+b_{m0}^+b_{mf}b_{nf} +
b_{nf}^+b_{mf}^+b_{m0}b_{n0}
\right\}
\endarray
$$
We have neglected  here the term
$$
\frac12{\sum\limits_{n,m}}'
\left(
\langle 00 \left| V_{nm} \right| 00 \rangle
-
\langle 0f \left| V_{nm} \right| 0f \rangle
+
\langle ff \left| V_{nm} \right| ff \rangle
\right)
b_{nf}^+b_{mf}^+b_{mf}b_{nf}
$$
which arises due to the multipolar interaction and is small comparing
with $H^3$ and $H^4$.
The matrix elements
$$
M_{nm}^f =
\langle 0f \left| V_{nm} \right| f0 \rangle,
$$
\equation
\label{simeq1}
\langle 00 \left| V_{nm} \right| ff \rangle
\simeq M_{nm}^f,
\endequation
\equation
\label{simeq2}
\langle ff \left| V_{nm} \right| 00 \rangle
\simeq M_{nm}^f,
\endequation
characterize the excitation `hopping'  from the site $n$ to $m$.
Introducing new (excitonic) operators

$$
P_{n} = b_{n0}^+b_{nf} \ , \qquad
P_{n}^+ = b_{nf}^+b_{n0}
$$
which play the role of the
creation ($P_{n}^+$) and annihilation ($P_{n}$) operators
of the exciton on the n-th
site, one can write the Hamiltonian in the finite form:

\eqnarray
{\bf H}^2 & = &
\sum\limits_{n} (\Delta\epsilon_f + {\cal D}_f) P_n^+P_n
\nonumber
\\
\label{exciton}
{\bf H}^3 & = & {\sum\limits_{n,m}}' M_{nm}^f P_m^+P_n
\\
\nonumber
{\bf H}^4 & = &  \frac12 {\sum\limits_{n,m}}' M_{nm}^f
\left(
P_m^+P_n^+ + P_m P_n
\right)
\endeqnarray

The operators $P_n$ form the Paulionic algebra, i.e. they satisfy the
following commutation relations.
They anticommute like fermionic operators on the same site:

\equation
\label{pauli1}
P_n  P_n^+  + P_n^+P_n = 1
\qquad  (P_n^+)^{2} = P_n^2=0
\endequation
and commute on the different sites like bosonic ones:
\equation
\label{pauli2}
P_n  P_m^{+} - P_m^+P_n = 0 \qquad
P_n^{+}  P_m^{+} - P_m^{+}P_n^{+} = 0 .
\endequation

In the second quantization representation the operator of dipole
momentum ${\bf p}_n$ looks like
$$
{\bf p}_n =
{\bf p}_n^{0f}
(b_{nf}^+b_{n0} + b_{n0}^+b_{nf})
$$
or, in excitonic operators
$$
{\bf p}_n = {\bf p}_n^{0f} (P_{n}^+ + P_{n})\, ,
$$
where ${\bf p}_n^{0f}$ is
the dipole momentum of the  0-$f$ transition.
In these terms the matrix elements $M_{nm}^f$ can be rewritten as:
$$
M_{nm}^f =
\frac{|{\bf p}_n^{0f}|^2 |{\bf n}-{\bf m}|^2
      - 3\bigl({\bf p}_n^{0f} \, , \,({\bf n}-{\bf m}) \bigr)^2}
     {|{\bf n}-{\bf m}|^5}\, .
$$
For the 1-D chain one can use the nearest
neighbor approximation, so in the following we shall
carry out summation in (\ref{exciton}) over $m=n\pm 1$ only:

$$
{\sum_{n,m}}'\longrightarrow\sum_{n, m=n\pm 1}
$$
This is strongly different from the 3-D situation when one can not
neglect distant mo\-le\-cu\-les, the number of which increases as
third degree of the distance, and therefore the term $\sum_{|n-m|>1}
\dots$ therefore has the same order as the nearest neighbors
contribution.

One should note that the coefficient $M_{nm}^f$ of the term $P_m^{+}
P_n^{+}$ is the same as that of the term $P_m^{+} P_n$ (see
(\ref{simeq1}, \ref{simeq2}) ) up to the effects of the
intermolecular electron exchange which we will not take into account.
This means, that {\it the problem of the optical response is
naturally resulted in XY-model rather than XX-model} which was
frequently treated earlier  \cite{Knoester,Jap}.

Now we will recall the formula for the linear response function.
Let
the aggregate be exposed in an external time dependent magnetic
field.
The energy of the aggregate  field interaction is equal to
that of the dipole momentum on the molecules in the electromagnetic
field:
$$
{\bf H}_t =
\sum_n
p_{n}^{0f} E_n(t)
\left(
P_n^+ + P_n
\right)
$$
(for every site the axis is chosen to be along with the molecular
dipole momentum direction).

In linear on the external field amplitude approximation the most
general relation between the statistical average of dipole momentum
$
\tilde{P_n} =
\langle
p_n^{0f} (P_n^+ + P_n)
\rangle
$
and the
electromagnetic field can be written as
$$
\tilde{P_n}(t) =
\sum_m
\int_0^\infty \mbox{d} \tau \
\sigma_{nm} (\tau)
E_m (t-\tau) \ ,
$$
where

\equation
\label{response}
\sigma_{nm} (t) = -
\frac{(p_{n}^{0f})^2}{ih}
\mathop{\mbox{Tr}}
\left(
[(P_m^+ + P_m),(P_n^+ + P_n)(t)]
\rho
\right)
\endequation
is the linear response function,
$\rho = \mbox{\large e}^{-\beta H}/\mathop{\mbox{Tr}}
\mbox{\large e}^{-\beta H}$ is the statistical operator,
$\beta$ is the inverse temperature.

Making use the identity
$$
\mathop{\mbox{Tr}}
\left(
\mbox{\large e}^{i{\bf H}t}
(P_m^+ + P_m)
\mbox{\large e}^{-i{\bf H}t}
(P_n^+ + P_n)
\rho
\right)
 =
\mathop{\mbox{ Tr}}
\left(
(P_m^+ + P_m)
\mbox{\large e}^{-i{\bf H}t}
(P_n^+ + P_n)
\mbox{\bf\large e}^{i{\bf H}t}
\rho
\right)
$$
one can write down expression (\ref{response}) in the form

\equation
\label{symmetric}
\sigma_{nm} (t) =
\frac{(p_n^{0f})^2}{ih}
\left(
\pi_{nm}(t) - \pi_{mn}(-t)
\right)
\endequation
where

$$
\pi_{nm}(t) =
\mathop{\mbox{Tr}}
\left(
(P_n^+ + P_n)(t)
(P_m^+ + P_m)
\rho
\right)
$$

Expanding the last expression we get the final formula for the
linear response function via the correlators of dipole momentum
operators:

$$
\sigma_{nm} (t) =
\frac{p_n^{0f}}{ih}
\left[
\langle
P_n^+(t)\  P_m^+
\rangle +
\langle
P_n^+(t)\  P_m
\rangle +
\langle
P_n(t)\  P_m^+
\rangle +
\langle
P_n(t)\  P_m
\rangle -
\right.
$$
\equation
\label{final}
\left.
- \langle
P_m^+(-t)\  P_n^+
\rangle -
\langle
P_m^+(-t)\  P_n
\rangle -
\langle
P_m(-t)\  P_n^+
\rangle -
\langle
P_m(-t)\  P_n
\rangle
\right]
\endequation
We can still simplify Eqn.(\ref{final}).
Indeed, due to the relation:

$$
\mathop{\mbox{Tr}}
\left(
\mbox{\large e}^{i{\bf H}t}
P_m
\mbox{\large e}^{-i{\bf H}t}
P_n
\mbox{\large e}^{-\beta {\bf H}}
\right)
=
\overline{
\mathop{\mbox{Tr}}
\left(
\mbox{\large e}^{-\beta {\bf H}}
P_n^+
\mbox{\large e}^{i{\bf H}t}
P_m^+
\mbox{\large e}^{-i{\bf H}t}
\right)
         }
$$
we have

$$
\left\langle
P_m(t)\ P_n
\right\rangle
=
\overline{
\left\langle
P_m^+\ P_n^+(t)
\right\rangle }
=
\overline{
\left\langle
P_n^+(-t)\ P_m^+
\right\rangle }
$$
Besides, the correlator
$
\left\langle
P_n^+(t)\ P_m
\right\rangle
$
is equal to
$
\left.
\left\langle
P_m(\tau)\ P_n^+
\right\rangle
\right|_{i\tau = \beta - it}
$
because of:

$$
\mathop{\mbox{Tr}}
\left(
\mbox{\large e}^{-(\beta-it){\bf H}}
P_n^+
\mbox{\large e}^{-i{\bf H}t}
P_m
\right)
=
\mathop{\mbox{Tr}}
\left(
\mbox{\large e}^{-i{\bf H}t}
P_m
\mbox{\large e}^{-(\beta-it){\bf H}}
P_n^+
\right)\, .
$$
Finally, using the time and the chain translation invariance
we derive the following expression for the response function
$\sigma _{nm}(t)$:

$$
\sigma _{nm}(t) =
\frac{2(p_n^{0f})^2}{h}
\biggl(
\mbox{Im}
\left\langle
P_n^+(t)\ P_m^+
\right\rangle +
\mbox{Im}
\left\langle
P_n^+(-t)\ P_m^+
\right\rangle +
$$
\equation
\label{*1}
\left.
+\mbox{Im}
\left\langle
P_n(t)\ P_m^+
\right\rangle +
\mbox{Im}
\left\langle
P_n(\tau)\ P_m^+
\right\rangle
\right|_{i\tau = \beta - it}
\biggr)\ .
\endequation
So, we have got the expression for the linear response function
$\sigma_{nm}(t)$.
Let us emphasize that
there are two different kinds of correlators in (\ref{*1}):
$
\left\langle
P_n^+(t)\ P_m^+
\right\rangle
$
and
$
\left\langle
P_n(t)\ P_m^+
\right\rangle
$.
In the next section these correlators will be found for the general
case of the cyclic XY-model.

\section{Correlators of XY-model}

In this section we will
consider the calculation of correlators which are
needed for the evaluation of the linear response following formula
(\ref{*1}).

The results of this section are valid for an
arbitrary XY-Hamiltonian
disregard to specific relations between its coefficients, so
we can write again Hamiltonian (\ref{sum}, \ref{exciton})
in the general notations:

\equation
\label{general}
{\bf H} =
\sum\limits_{n=1}^M
\epsilon P_n^+P_n +
h \sum_{n=1}^{M}\sum_{m=n\pm 1} P_m^+P_n +
\frac12
\gamma
\sum_{n=1}^{M}\sum_{m=n\pm 1}
\left(
P_m^+P_n^+ + P_m P_n
\right)
\endequation
where the cyclic boundary conditions are imposed and
$P_n^+$, $P_n$ are paulionic creation and annihilation
operators obeying
the commutation relations (\ref{pauli1}, \ref{pauli2}).
For excitons the parameters are
$\epsilon \equiv \Delta\epsilon_f + {\cal D}_f$,
$h=\gamma\equiv M_{nm}^f$
We have
also eliminated the constant summand ${\bf H}^1$ (\ref{H1}),
because statistical characteristics are independent
on a constant shift in the energy operator.

It is convenient to start with the consideration of the
partition function of the system.
After that the correlators
$\langle P_1(t)\ P_{L+1}^+(0)\rangle $ and
$\langle P_1^+(t)\ P_{L+1}^+(0)\rangle $ will be evaluated.

\subsection{Partition function}

As usually, we make use the Jordan--Wigner transformation

$$
P_m =  (-1)^{\sum\limits^{m-1}_{k=1} c_k^+ c_k} c_m \ ,
\qquad
P_m^+ = c_m^+ (-1)^{\sum\limits^{m-1}_{k=1} c_k^+ c_k}
$$
to rewrite operator (\ref{general}) in the fermionic Hamiltonian:

\equation
\label{fermi}
{\bf H} =
\epsilon\sum\limits_{n=1}^M
c_n^+c_n +
h\sum_{n=1}^M
\left(
c_n^+c_{n+1} + c_{n+1}^+c_n
\right) +
\gamma
\sum_{n=1}^M
\left(
c_n^+c_{n+1}^+ + c_{n+1} c_n
\right)
\endequation
with the following boundary conditions for the fermionic creation
and annihilation operators:

\begin{equation}
c_{M+1} = \mbox{\large e}^{-i\pi\hat N} c_1 \qquad
c^+_{M+1} = c^+_1 \mbox{\large e}^{i\pi\hat N} \qquad
\hat N=\sum\limits^M_{n=1} c^+_n c_n    \ .
\end{equation}
Let us denote the orthogonal projection operators on states
containing even and odd number of particles by $\mbox{P}_+$ and
$\mbox{P}_-$ respectively:

\equation
\label{ortho}
\mbox{P}_\pm =
\frac12
\left(
1\pm (-1)^{\hat{N}}
\right)
\endequation
and consider the operators ${\bf H}_\pm$  of form (\ref{fermi}) with
the boundary conditions

$$
c_{M+1} = \mp c_1 \qquad
c^+_{M+1} =\mp c^+_1 \ .
$$
In the momentum
representation
$$
c_m = \frac{e^{\pi/4}}{\sqrt M} \sum\limits_{p\in A_\pm}
\mbox{\large e}^{ipm} c_p ,\qquad
A_\pm =
\left\{
p\in (-\pi,\pi] : \mbox{\large e}^{ipM} = \mp 1
\right\}
$$
the operators ${\bf H}_\pm$ take the quasidiagonal form

$$
{\bf H}_\pm = \sum\limits_{p\in A_\pm}
\left\{
\varepsilon_p c^+_p c_p +
\gamma
\sin p
(c^+_p c^+_{-p} + c_{-p} c_p)
\right\}
\ ,
\qquad
\varepsilon_p = \epsilon + 2h\cos(p)
$$
Then the partition function $Z$ with the chemical
potential $\mu$ is represented as:

$$
Z=
\mbox{Tr}
\left\{
\exp (-\beta({\bf H}_+ -\mu N)) \mbox{P}_+
\right\}
+
\left\{
\mbox{Tr} \exp (-\beta({\bf H}_- -\mu N)) \mbox{P}_-
\right\}
$$
that can be rewritten in the form:

\eqnarray
\nonumber
Z & = &
 \frac12  \left(
              \mathop{\mbox{Tr}}
              \mbox{\large e}^{-\beta({\bf H}_+-\mu\hat{N})} +
              \mathop{\mbox{Tr}}
              \mbox{\large e}^{-\beta({\bf H}_--\mu\hat{N})} +
\right.\\
\label{4}
          && +  \left. \mathop{\mbox{Tr}}(-1)^{\hat{N}}
              \mbox{\large e}^{-\beta({\bf H}_+-\mu\hat{N})} -
              \mathop{\mbox{Tr}}(-1)^{\hat{N}}
              \mbox{\large e}^{-\beta({\bf H}_--\mu\hat{N})}
              \right)
\endeqnarray
Such form is  convenient for the calculation of the partition
function.  Indeed, each term of Eqn.(\ref{4}) is a partition function
of a quadratic Hamiltonian which is quasidiagonal in the momentum
representation.  Moreover, the moments $k$ and $-k$ are only
connected.  So evaluating traces in each block and taking the product
over all momenta we have the following result for the partition
function \cite{IKK}:

\equation
\label{Z}
Z =
\frac{1}{2}
(Z_{f}^{+} + Z_{f}^{-} + 1/Z_{b}^{+} - 1/Z^{-}_{b}) \ .
\endequation
Here $Z_{f}^{\pm}$ $(Z_{b}^{\pm})$ are the fermionic (bosonic)
partition functions

\eqnarray
\nonumber
Z_{f}^{\pm} &= &
\mbox{\large e}^{-\beta\epsilon M/2}
\prod_{k_{\pm}}
\mbox{\large e}^{\beta E(k_\pm)/2}
\left(
1+\mbox{\large e}^{-\beta E(k_\pm)}
\right)
\\
\nonumber
Z_{b}^{\pm} &= &
\mbox{\large e}^{\beta \epsilon M/2}
\prod_{k_{\pm}}
\mbox{\large e}^{-\beta E(k_\pm)/2}
\left(
1-\mbox{\large e}^{-\beta E(k_\pm)}
\right)^{-1}
\endeqnarray
for systems with the energy spectra

$$
E(k_{\pm}) =
\left(
\varepsilon_{k_\pm}-\mu
\right)
\sqrt{
1 +
\frac{4\gamma^2\sin^2(k_\pm)}
     {(\epsilon_{k_\pm}-\mu)^2}
     }
$$
and the antiperiodic (periodic) boundary conditions:
$$
k_{+}=\frac{2\pi}{M} (m+1/2)\ ,
\quad
k_{-}=\frac{2\pi}{M}m  \ ,
\qquad
m=0,\ldots ,M-1 \ .
$$
Using formula (\ref{Z}) one can calculate all equilibrium
thermodynamical quantities.  The partition function will be used
further for calculations of the correlators. There we will put
$\mu = 0$.

\subsection{Correlator $\langle P_1(t)\ P_{L+1}^+(0)\rangle $}

In this subsection we concentrate our attention
on the correlator
$\langle P_1(t)\ P_{L+1}^+(0)\rangle $.
It will be calculated by the coherent states method adopted for the
case of the cyclic XY-model in Ref.\cite{KI}
(the final result of the paper contains a mistake corrected in
Ref.\cite{IKK}).

By the definition, the correlator we are interested in is

\equation
\label{Tr}
\langle P_1(t)\ P_{L+1}^+(0)\rangle =
\frac1Z
\mbox{Tr}
\left(
\mbox{\large e}^{-(\beta - it){\bf H}}
c_1
\mbox{\large e}^{-i{\bf H}t}
c_{L+1}^+
(-1)^{\sum_{n=1}^L c_n^+ c_n}
\right)
\endequation
As in the previous section it allows us to rewrite correlator
(\ref{Tr}) in the following form:

\equation
\label{fb}
\langle P_1(t)\ P_{L+1}^+(0)\rangle =
\frac{1}{2Z}
\left\{
K_f^+ +
K_b^+ +
K_f^- -
K_b^-
\right\}
\endequation
where we have introduced the notations:

\eqnarray
\nonumber
K_f^\pm &=&
\mbox{Tr}
\left(
\mbox{\large e}^{-(\beta - it){\bf H}_\pm}
c_1
\mbox{\large e}^{-i{\bf H}_\mp t}
c_{L+1}^+
(-1)^{\sum_{n=1}^L c_n^+ c_n}
\right)
\\
\nonumber
K_b^\pm &=&
\mbox{Tr}
\left(
\mbox{\large e}^{-(\beta - it){\bf H}_\pm}
c_1
\mbox{\large e}^{-i{\bf H}_\mp t}
c_{L+1}^+
(-1)^{\sum_{n=L+1}^M c_n^+ c_n}
\right)
\endeqnarray

Let us calculate accurately the first term $K_f^+$.
We will consider the basis of coherent states

$$
\left|
\xi_{p_\pm}
\right\rangle =
\mbox{\large e}^{\sum_{p_\pm} c_{p_\pm}^+ \xi_{p_\pm}}
\left| 0 \right\rangle \
$$
with the following well-known properties:

\eqnarray
\nonumber
\mbox{Tr} A &= &
\int \,
\prod_{p_\pm} \mbox{d} \xi_{p_\pm}\, \mbox{d} \bar{\xi}_{p_\pm}\
\mbox{\large e}^{\bar{\xi} \xi }
\left\langle
\bar{\xi}_{p_\pm}
\right|
A
\left|
\xi_{p_\pm}
\right\rangle
\\
\nonumber
\mbox{id} &=&
\int
\prod_{p_\pm} \mbox{d} \bar{\xi}_{p_\pm}\, \mbox{d} \xi_{p_\pm}\
\mbox{\large e}^{-\bar{\xi}\xi }
|
\xi_{p_\pm}
\rangle
\langle
\bar{\xi}_{p_\pm}
|
\endeqnarray
The variables $\xi$ are Grassmann numbers and are taken to be
commuting with  operators.  In the Bargmann-Fock representation
correlator $K_f^+$ is expressed as follows

\eqnarray
\nonumber
K_f^+ &=&
\int
\prod_{p_+} \mbox{d} \xi_{p_+}^0\, \mbox{d} \bar{\xi}_{p_+}^0\,
\int
\prod_{p_+} \mbox{d} \bar{\xi}_{p_+}^1\, \mbox{d} \xi_{p_+}^1\,
\int
\prod_{p_+} \mbox{d} \bar{\xi}_{p_+}^2\, \mbox{d} \xi_{p_+}^2 \times
\nonumber\\
&&\times\exp
\sum_{p_+}
\left(
\bar{\xi}_{p_+}^0\xi_{p_+}^0 -
\bar{\xi}_{p_+}^1\xi_{p_+}^1 -
\bar{\xi}_{p_+}^2\xi_{p_+}^2
\right)
\left\langle
\bar{\xi}_{p_+}^0
\right|
\mbox{\large e}^{-(\beta - it){\bf H}^+} c_1
\left|
\xi_{p_+}^1
\right\rangle
\times
\nonumber
\\
\label{Kf}
& &\times
\left\langle
\bar{\xi}_{p_+}^1
\right|
\mbox{\large e}^{- it{\bf H}^-}
\left|
\xi_{p_+}^2
\right\rangle
\left\langle
\bar{\xi}_{p_+}^2
\right|
c_{L+1}^+
(-1)^{\sum_{k=1}^{L}c_k^+c_k}
\left|
\xi_{p_+}^0
\right\rangle
\endeqnarray
We will consider each multiplier in details and start with the third
one.  Making use the identity (N is the normal ordering symbol)
$$
\mbox{\large e}^{\alpha c^+ c} =
\mbox{N}\exp
\left(
(\mbox{\large e}^\alpha - 1)c^+c
\right)
$$
one can rewrite the operator
$(-1)^{\sum_{k=1}^L c_k^+c_k}$ in the normal ordered form:

$$
(-1)^{\sum_{k=1}^L c_k^+c_k} =
\mbox{N}\exp
\left\{
-2\sum_{p_+,p_+'}
c_{p_+}^+c_{p_+'}
V_{p_+p_+'}^f
\right\}
$$
with the matrix $V^f$:
\equation
V_{p_+p_+'}^f =
\frac1M
\frac{\sin\frac{1}{2} (p_+' -p_+)L}
     {\sin\frac{1}{2} (p_+'  -p_+)}\,
\mbox{\large e}^{\frac12 i(p_+' -p_+)(L+1)}\ .
\label{V}
\endequation
Then, substituting the momentum representation for the operator
$c_{L+1}^+$

\equation
\label{cp}
c_{L+1}^+ =
\frac{\mbox{\large e}^{-i\pi/4}}{\sqrt{M}}
\sum_{p_+}
\mbox{\large e}^{-ip_+(L+1)}
c_{p_+}^+
\endequation
we come to the expression for the third multiplier in (\ref{Kf}):

\equation
\label{3term}
\left\langle
\bar{\xi}_{p_+}^2
\right|
c_{L+1}^+
(-1)^{\sum_{k=1}^{L}c_k^+c_k}
\left|
\xi_{p_+}^0
\right\rangle
=
\frac{\mbox{\large e}^{-i\pi/4}}{\sqrt{M}}
\sum_{p_+''}
\mbox{\large e}^{-ip_+''(L+1)}
\bar{\xi}_{p_+''}^2
\exp \left\{-\sum_{p_+,p_+'}
           \bar{\xi}_{p_+}^2
           W_{p_+p_+'}^f\xi_{p_+'}^0  \right\}
\endequation
where

$$
W_{p_+,p_+'}^f =
-2V_{p_+p_+'}^f + \delta_{p_+p_+'}\ .
$$

Let us now consider the first multiplier in the integral (\ref{Kf}).
Using (\ref{cp}) one can rewrite it as follows:

\equation
\label{first}
\left\langle
\bar{\xi}_{p_+}^0
|
\mbox{\large e}^{-(\beta - it){\bf H}_+} c_1
|
{\xi}_{p_+}^1
\right\rangle  =
\frac{\mbox{\large e}^{i\pi/4}}{\sqrt{M}}
\sum_{p_{+}^{\prime\prime\prime}}
\mbox{\large e}^{ip_{+}^{\prime\prime\prime}}
\xi^{1}_{p_{+}^{\prime\prime\prime}}
\left\langle
\bar{\xi}_{p_+}^0
|
\mbox{\large e}^{-(\beta - it){\bf H}_+}
|
-{\xi}_{p_+}^1
\right\rangle
\endequation
To calculate the matrix element we have to present the operator
$\mbox{\large e}^{-(\beta - it){\bf H}_+}$ in the normal ordered
form.  First of all we will extract terms in ${\bf H}_{+}$ with the
momentum $p=\pi$ (if it is presented at all).  All other terms can be
grouped in pairs with opposite momenta, $\{p_{+},-p_{+}\}$.  Then the
exponent can be put down:

\equation
\label{pi}
\mbox{\large e}^{-(\beta - it){\bf H}_+} =
\mbox{\large e}^{-(\beta - it){\bf H}_+^\pi}
\prod_{\{p_+,-p_+\}\ne \pi}
\mbox{\large e}^{-(\beta - it){\bf H}_+^{p_+}}
\endequation
where the first multiplier is only presented if $M$ is odd and

$$
{\bf H}_+^p =
\varepsilon_p
\left(
c_p^+ c_p + c_{-p}^+ c_{-p}
\right)+
2\gamma\sin p
\left(
c_{-p} c_p +  c_p^+c_{-p}^+
\right)
$$
One can derive the following expression for the normal ordered
exponent of the last operator:

\eqnarray
\nonumber
\mbox{\large e}^{-(\beta - it){\bf H}^{p_+}_+}  &=&
\mbox{\large e}^{-(\beta - it)\varepsilon_{p_+}}
\,
K_{p_+}(\beta - it)
\mbox{N}\exp
\Biggl\{
\left(
\frac{1}{K_{p_+}(\beta - it)} - 1
\right)
(c_{p_+}^+ c_{p_+} + c_{-p_+}^+ c_{-p_+})+
\\
& &
\nonumber
+ \sigma_{p_+}(\beta - it)
\left(
c_{p_+}^+ c_{-p_+}^+ +
c_{-p_+} c_{p_+}
\right)
\Biggr\}
\endeqnarray
where

\eqnarray
\nonumber
K_p(\beta) &=&
\cos^2 \frac{\theta_p}{2}
\mbox{\large e}^{\beta E_p}
+
\sin^2 \frac{\theta_p}{2}
\mbox{\large e}^{-\beta E_p}
\\
\nonumber
\sigma_p(\beta) &= &
\frac{1}{K_p(\beta)} \sin\theta_p
\mbox{sh}\beta E_p
\endeqnarray
\nonumber
and $\theta_p$ is so-called Bogoliubov's angle:
$$
\sin \theta_{p} =
-\frac{2\gamma \sin p}{E_{p}} \ ,
\qquad
\cos \theta_{p} =
\frac{\varepsilon_{p}}{E_{p}} \ ,
\qquad
E_{p} =
\varepsilon_{p}
\sqrt{\frac{4\gamma^{2}\sin^{2}p}{\varepsilon^{2}_{p}} + 1} \ .
$$
Now one can get the expression for the matrix element in
(\ref{first}):

\eqnarray
\nonumber
&&
\left\langle
\bar{\xi}_{p_+}^0
\right|
\mbox{\large e}^{- (\beta-it){\bf H}_+}
\left|
-\xi_{p_+}^1
\right\rangle =
\prod_{\{ p_+,-p_+\}\ne\pi}
\mbox{\large e}^{-(\beta - it)\varepsilon_{p_+}}
K_{p_+}(\beta - it)\times
\\
&&
\nonumber
\qquad\times
\exp
\Biggl\{
- \frac{1}{K_{p_+}(\beta - it)}
\left(
\bar{\xi}^{0}_{ p_{+}} \xi^{1}_{ p_{+}} +
\bar{\xi}^{0}_{-p_{+}} \xi^{1}_{-p_{+}}
\right)  -
\\
\nonumber
& &
-\sigma_{p_+}(\beta - it)
\left(
\bar{\xi}^{0}_{p_{+}} \bar{\xi}^{0}_{-p_{+}} +
\xi^{1}_{-p_{+}}\xi^{1}_{p_{+}}
\right)
\Biggr\}
\exp
\left(
-\mbox{\large e}^{-(\beta - it)\varepsilon_\pi}
\bar{\xi}^{0}_\pi \xi^{1}_\pi
\right)
\endeqnarray
This expression allows us to  consider the product over all momenta
$p_{+}$ (instead of the product over the pairs $\{p_{+},-p_{+}\}$).
Then the terms with the momentum $k=\pi$ can be taken into account
in the product over all $p_{+}$ and the term (\ref{first})
can be presented as

\eqnarray
 & &
\nonumber
\left\langle
\bar{\xi}_{p_+}^0
\right|
\mbox{\large e}^{- (\beta-it){\bf H}_+} c_{1}
\left|
\xi_{p_+}^1
\right\rangle
=
\frac{\mbox{\large e}^{i\pi/4}}{\sqrt{M}}
\sum_{p_{+}'}
\mbox{\large e}^{ip_{+}'}\
\xi^{1}_{p_{+}'}
\left(
\prod_{p_+}
\mbox{\large e}^{-(\beta - it)\varepsilon_{p_+}}
K_{p_+}(\beta-it)
\right)^{\frac12} \times
\\
\label{1term}
& &
\qquad\times
\exp
\left\{
       -\frac{1}{K_{p_+}(\beta - it)}
        \bar{\xi}^{0}_{p_{+}} \xi^{1}_{p_{+}}  -
        \frac{\sigma_{p_+}(\beta-it) }{2}
        \left(
          \bar{\xi}^{0}_{p_{+}}
          \bar{\xi}^{0}_{-p_{+}} +
          \xi^{1}_{-p_{+}}
          \xi^{1}_{p_{+}}
        \right)
\right\}
\endeqnarray

In close analogue with this term we can derive an expression for the
second matrix element in (\ref{Kf}).  It is given by

\eqnarray
\nonumber
\left\langle
\bar{\xi}_{p_+}^1
\right|
\mbox{\large e}^{-it{\bf H}^-} c_{1}
\left|
\xi_{p_+}^2
\right\rangle &=&
\left(
\prod_{p_-}
\mbox{\large e}^{- it\varepsilon_{p_-}}
K_{p_-}(it)
\right)^{\frac12}
\times
\\
& &
\label{2term}
\times
\exp
\left\{
  \frac{1}{K_{p_-}(it)}
  \bar{\xi}^{1}_{p_{-}}
  \xi^{2}_{p_{-}}
  - \frac{\sigma_{p_-}(it)}{2}
  \left(
    \bar{\xi}^{1}_{p_{-}}
    \bar{\xi}^{1}_{-p_{-}} +
    \xi^{2}_{-p_{-}}
    \xi^{2}_{p_{-}}
  \right)
\right\}
\endeqnarray
where the variables $\xi_{p_-}$ are connected with the variables
$\xi_{p_+}$ by the unitary transformation:

$$
\xi_{p_{-}} =
\sum_{p_{+}} U_{p_{-}p_{+}} \xi_{p_{+}} \ ,
\qquad
\bar{\xi}_{p_{-}} =
\sum_{p_{+}} \bar{\xi}_{p_{+}}\ U^{+}_{p_{+}p_{-}} \ ,
\qquad
U_{p_{-}p_{+}} =
\frac{1}{M}
\sum_{k=1}^{M}
\mbox{\large e}^{ik(p_{+}-p_{-})} \ .
$$
Substituting (\ref{3term}), (\ref{1term}) and (\ref{2term}) in the
integral (\ref{Kf}) one can get the final integral expression:

\eqnarray
\nonumber
K_{f(b)}^+ &= &
\left(
        \prod_{p_+}
        \mbox{\large e}^{-(\beta - it)\varepsilon_{p_+}}
        K_{p_+}(\beta-it)
        \prod_{p_-}
        \mbox{\large e}^{- it\varepsilon_{p_-}}
        K_{p_-}(it)
\right)^{\frac12}
\frac1M
\sum_{p_+,p_{+}'}
\mbox{\large e}^{ip_+' - ip_+(L+1)}
\times
\\
& &
\times
\label{K+}
\int
\mbox{d} \xi^0\, \mbox{d} \bar{\xi}^0\,
\mbox{d} \bar{\xi}^1\,\mbox{d} \xi^1\,
\mbox{d} \bar{\xi}^2\,\mbox{d} \xi^2 \
\exp
\left\{
\bar{\xi}^0\xi^0 -
\bar{\xi}^1\xi^1 -
\bar{\xi}^2\xi^2
\right\} \,
\xi_{p_+'}^1\bar{\xi}_{p_+}^2
\times
\\
& &
\nonumber
\times
\exp
\left\{
-\bar{\xi}^0D_+^{(1)}\xi^1 +
\right.
+ \bar{\xi}^0B_+^{(1)}\bar{\xi}^0 -
\xi^1B_+^{(1)}\xi^1 +
\bar{\xi}^1U^+D_-^{(2)}U\xi^2 +
\\
& &
\nonumber
\left.
+\bar{\xi}^1U^+B_-^{(2)}\bar{U}\bar{\xi}^1 -
\xi^2U^tB_-^{(2)}U\xi^2 -
\bar{\xi}^2W_+^{f(b)}\xi^0
\right\}
\endeqnarray
where we use the matrix notations:

\eqnarray
\nonumber
D^{(1)} &= \mbox{diag} (\frac{1}{K_p(\beta - it)})\ ,
\qquad
B^{(1)}_{pp'} &=
-\delta_{p,-p'}
\frac{\sigma_p(\beta - it)}{2}
\\
\nonumber
D^{(2)} &= \mbox{diag} (\frac{1}{K_p(it)})\ ,
\qquad
B^{(2)}_{pp'} &=
-\delta_{p,-p'}
\frac{\sigma_p(it)}{2}
\endeqnarray

indices ``$+$'' and ``$-$'' at matrices point the momentum space
($A_{+}$ or $A_{-}$) in which these indices are considered.

One can see that $K_{b}^{+}$ can be written in the same form as
$K_{f}^{+}$.  There is the only difference in the third matrix
element in (\ref{Kf}).  It leads to replacing of the matrix
$W^{f}_{+}$ by the matrix $W_{+}^{b}$.  The matrix $W_{+}^{b}$ can be
easy calculated and results in
$$
W_{p_+p_+'}^b =
\delta_{p_+p_+'}
- \frac2M
\exp
\left\{
\frac12 i(p_+' -p_+)(M+L+1)
\right\}
\frac{ \sin\frac12 (p_+' - p_+)(M-L) }
     { \sin\frac12 (p_+' - p_+)      } =
-W_{p_+p_+'}^f
$$
That is why we have the general expression for $K_{f}^{+}$ and
$K_{b}^{+}$ in (\ref{K+}).

To obtain $K_{f(b)}^{-}$ one should replace the matrix $U$ by the
matrix $U^{+}$, ``$+$'' by ``$-$'' (and visa-versa) in
Eqn.(\ref{K+}).  It can be easy understood from the calculating
procedure.  Thus $K^{-}_{f(b)}$ is presented by:

\eqnarray
\nonumber
K_{f(b)}^- &= &
\left(
\prod_{p_-}
\mbox{\large e}^{-(\beta - it)\varepsilon_{p_-}}
K_{p_-}(\beta - it)
\prod_{p_+}
\mbox{\large e}^{- it\varepsilon_{p_+}}
K_{p_+}(it)
\right)^{\frac12}
\frac1M
\sum_{p_-,p_{-}'}
\mbox{\large e}^{ip_-' - ip_-(L+1)}
\times
\\
& &
\nonumber
\times
\int
\mbox{d} \xi^0\, \mbox{d} \bar{\xi}^0\,
\mbox{d} \bar{\xi}^1\,\mbox{d} \xi^1 \,
\mbox{d} \bar{\xi}^2\, \mbox{d} \xi^2\
\exp
\left\{
\bar{\xi}^0\xi^0 -
\bar{\xi}^1\xi^1 -
\bar{\xi}^2\xi^2
\right\}
\xi_{p_-'}^1\bar{\xi}_{p_-}^2
\times
\\
& &
\nonumber
\times
\exp
\left\{
-\bar{\xi}^0 D_-^{(1)}\xi^1 +
\right.
+ \bar{\xi}^0 B_-^{(1)}\bar{\xi}^0 -
 \xi^1B_-^{(1)}\xi^1 +
\bar{\xi}^1UD_+^{(2)}U^+\xi^2 +
\\
& &
\left.
+\bar{\xi}^1UB_+^{(2)}U^t\bar{\xi}^1 -
\xi^2\bar{U}B_+^{(2)}U^+\xi^2 -
\bar{\xi}^2W_-^{f(b)}\xi^0
\right\}
\label{K-}
\endeqnarray
So we have obtained all terms in (\ref{fb}).  Now we should calculate
the integrals in expressions (\ref{K+},\ref{K-}).  To do it we
introduce four generating functions $I^{+}_{f(b)}(\bar{\eta},\eta)$
and $I^{-}_{f(b)}(\bar{\eta},\eta)$ as follows:

\eqnarray
\nonumber
I^{+}_{f(b)}(\bar{\eta},\eta) &=&
\int
\mbox{d} \xi^0\, \mbox{d} \bar{\xi}^0\,
\mbox{d} \bar{\xi}^1\,\mbox{d} \xi^1\,
\mbox{d} \bar{\xi}^2\, \mbox{d} \xi^2\
\exp
\left(
\bar{\xi}^0\xi^0 -
\bar{\xi}^1\xi^1 -
\bar{\xi}^2\xi^2
\right)
\times
\\
& &
\nonumber
\times
\exp
\left\{
-\bar{\xi}^0D_+^{(1)}\xi^1 +
\bar{\xi}^0B_+^{(1)}\bar{\xi}^0 -
\xi^1B_+^{(1)}\xi^1 +
\bar{\xi}^1U^+D_-^{(2)}U\xi^2 +
\right.
\\
\nonumber
& &
\left.
+\bar{\xi}^1U^+B_-^{(2)}\bar{U}\bar{\xi}^1 -
\xi^2U^tB_-^{(2)}U\xi^2 -
\bar{\xi}^2W_+^{f(b)}\xi^0 +
\bar\eta\xi^1 +
\bar{\xi}^2\eta
\right\}     \ .
\endeqnarray

\eqnarray
I^{-}_{f(b)}(\bar{\eta},\eta) &=&
\int
\mbox{d} \xi^0\, \mbox{d} \bar{\xi}^0\,
\mbox{d} \bar{\xi}^1\,\mbox{d} \xi^1 \,
\mbox{d} \bar{\xi}^2\,\mbox{d} \xi^2 \
\exp
\left(
\bar{\xi}^0\xi^0 -
\bar{\xi}^1\xi^1 -
\bar{\xi}^2\xi^2
\right)
\times
\nonumber
\\
& &
\label{I-}
\times
\exp
\left\{
-\bar{\xi}^0D_-^{(1)}\xi^1 +
\bar{\xi}^0B_-^{(1)}\bar{\xi}^0 -
\xi^1B_-^{(1)}\xi^1 +
\bar{\xi}^1UD_+^{(2)}U^+\xi^2 +
\right.
\\
\nonumber
& &
\left.
+\bar{\xi}^1UB_+^{(2)}U^t\bar{\xi}^1 -
\xi^2\bar{U}B_+^{(2)}U^+\xi^2 -
\bar{\xi}^2W_-^{f(b)}\xi^0 +
\bar\eta\xi^1 +
\bar{\xi}^2\eta
\right\} \ .
\endeqnarray
The detaled calculation of these integrals  and corresponding
derivatives is placed in Appendix A.
Here we only put down the final results for $K_{f(b)}^\pm$:

\eqnarray
\nonumber
K^\pm_{f(b)} &=&
(-1)^{\frac{M(M-1)}{2}}
\mathop{\mbox{Pf}}(G_\pm^{f(b)})
\sqrt{\det R_{\mp}^{-1}}
\times
\\
\nonumber
& &
\times
\left(
\prod_{p_\pm}
\mbox{\large e}^{- (\beta-it)\varepsilon_{p_\pm}}
K_{p_\pm}(\beta - it)
\prod_{p_\mp}
\mbox{\large e}^{-it\varepsilon_{p_\mp}}
K_{p_\mp}(it)
\right)^{\frac12}
\times
\\
& &
\label{**}
\times
\frac1M
\sum_{p_\pm ,p_\pm'}
\mbox{\large e}^{ip_\pm' - ip_\pm (L+1)}
        \left(
                Q_\pm^{f(b)} +
                T_\pm^{f(b)}
                \left(
                        Q_\pm^{f(b)}
                \right)^{-t}
                S_\pm
        \right)_{p_\pm'p_\pm}^{-1}
\endeqnarray
with the following notations:

\eqnarray
Q_+^{f(b)} & = & W_+^{f(b)}D_+^{(1)} + U^+ D_-^{(2)} R_-U
\nonumber
\\
\label{QTS}
T_+^{f(b)} & = &
2\left\{
W_+^{f(b)}B_+^{(1)}\left( W_+^{f(b)}\right)^t +
U^+ B_-^{(2)}R_-\bar{U}
\right\}
\\
\nonumber
S_+ & = &
-2\left\{
B_+^{(1)} +
U^t B_-^{(2)} R_- U
\right\}
\endeqnarray

\eqnarray
Q_-^{f(b)} & = & W_-^{f(b)}D_-^{(1)} + U D_+^{(2)} R_+ U^+
\nonumber
\\
\label{QTS-}
T_-^{f(b)} & = &
2\left\{
W_-^{f(b)}B_-^{(1)}\left( W_-^{f(b)}\right)^t +
U B_+^{(2)} R_+ U^t
\right\}
\\
\nonumber
S_- & = &
-2\left\{
B_-^{(1)} +
\bar{U} B_+^{(2)} R_+ U^+
\right\}
\endeqnarray
and the matrix $G_{\pm}^{f(b)}$ has the form
\equation
G_{\pm}^{f(b)} =
\left(
\array{cc}
S_{\pm} & \left( Q_{\pm}^{f(b)}\right)^t
\\
- Q_{\pm}^{f(b)} & T_{\pm}^{f(b)}
\endarray
\right)    \ ,
\endequation
where the matrix $R$ is defined as

\equation
R =
\left\{  \left( D^{(2)} \right)^2 - 4 \left( B^{(2)} \right)^2
\right\}^{-1} =
\mbox{diag}
\left\{
\frac{K_{p}^2(it)}{1+K_{p}^2(it)\sigma_{p}^2(it)}
\right\}               \ .
\label{R}
\endequation

\subsection{Correlator
$
\left\langle
P_1^+(t)\ P_{L+1}^+(0)
\right\rangle
$
}

Let us now consider another correlator

$$
\left\langle
P_1^+(t)\ P_{L+1}^+(0)
\right\rangle
\equiv
\frac1Z\mbox{Tr}
\left(
\mbox{\large e}^{-(\beta-it){\bf H}}
c_1^+
\mbox{\large e}^{-it{\bf H}}
c_{L+1}^+
(-1)^{\sum_{k=1}^L c_k^+c_k}
\right)
$$
In close analogue with
$
\left\langle
P_1(t)\ P_{L+1}^+(0)
\right\rangle
$
one can rewrite the correlator as:

\equation
\left\langle
P_1^+(t)\ P_{L+1}^+(0)
\right\rangle
=
\frac{1}{2Z}
\left(
\tilde{K}_f^+ +
\tilde{K}_b^+ +
\tilde{K}_f^- -
\tilde{K}_b^-
\right)
\label{I++}
\endequation
where

\eqnarray
\nonumber
\tilde{K}_f^\pm &=&
\mbox{Tr}
\left(
\mbox{\large e}^{-(\beta-it){\bf H}_\pm}
c_1^+
\mbox{\large e}^{-it{\bf H}_\mp}
c_{L+1}^+
(-1)^{\sum_{k=1}^L c_k^+c_k}
\right)
\\
\nonumber
\tilde{K}_b^\pm &=&
\mbox{Tr}
\left(
\mbox{\large e}^{-(\beta-it){\bf H}_\pm}
c_1^+
\mbox{\large e}^{-it{\bf H}_\mp}
c_{L+1}^+
(-1)^{\sum_{k=L+1}^M c_k^+c_k}
\right)
\endeqnarray

The calculating procedure is the same as in the previous subsection.
But there are several differences on which we will concentrate here.

First of all, let us consider expression (\ref{Kf}).  There is the
only difference here for $\tilde{K}^{+}_{f}$ in two first matrix
elements in (\ref{Kf}). For $\tilde{K}^{+}_{f}$ they are:

$$
\left\langle
\bar{\xi}_{p_+}^0
\left|
\mbox{\large e}^{-(\beta-it){\bf H}_+}
\right|
\xi_{p_+}^1
\right\rangle
\left\langle
\bar{\xi}_{p_+}^1
\left|
c_1^+
\mbox{\large e}^{-it{\bf H}_-}
\right|
\xi_{p_+}^2
\right\rangle
$$
to provide the normal ordering.  We have estimated all matrix element
for $K^{+}_{f}$ in the previous section in details.  From these
calculations one can see that for $\tilde{K}^{+}_{f}$ we should
change the matrix $D$ by the matrix $-D$ in (\ref{K+}), change the
sign in one of the exponents in front of the integral and put the
multiplier $\mbox{\large e}^{-i\pi/2}=-i$ due to the properties of
Fourier transformation.

So one can infer the expression for $\tilde{K}^{+}_{f(b)}$:

\eqnarray
\nonumber
\tilde{K}_{f(b)}^+ &= &
\left(
        \prod_{p_+}
        \mbox{\large e}^{-(\beta - it)\varepsilon_{p_+}}
        K_{p_+}(\beta - it)
        \prod_{p_-}
        \mbox{\large e}^{- it\varepsilon_{p_-}}
        K_{p_-}(it)
\right)^{\frac12}
\frac{(-i)}{M}
\sum_{p_+,p_{+}'}
\mbox{\large e}^{- ip_+' - ip_+(L+1)}
\times
\\
& &
\nonumber
\times
\int
\mbox{d} \xi^0 \,\mbox{d} \bar{\xi}^0\,
\mbox{d} \bar{\xi}^1\,\mbox{d} \xi^1\,
\mbox{d} \bar{\xi}^2\,\mbox{d} \xi^2 \
\exp
\left\{
\bar{\xi}^0\xi^0 -
\bar{\xi}^1\xi^1 -
\bar{\xi}^2\xi^2
\right\}
\bar{\xi}_{p_+}^1\bar{\xi}_{p_+'}^2
\times
\\
& &
\nonumber
\times
\exp
\left\{
\bar{\xi}^0D_+^{(1)}\xi^1 +
\right.
+ \bar{\xi}^0B_+^{(1)}\bar{\xi}^0 -
\xi^1B_+^{(1)}\xi^1 -
\bar{\xi}^1U^+D_-^{(2)}U\xi^2 +
\\
& &
\label{fK+}
\left.
+\bar{\xi}^1U^+B_-^{(2)}\bar{U}\bar{\xi}^1 -
\xi^2U^tB_-^{(2)}U\xi^2 -
\bar{\xi}^2W_+^{f(b)}\xi^0
\right\}
\endeqnarray
where all notations have been introduced in the previous subsection.
Using the same arguments as above for $K_{f(b)}^{-}$ we can derive
the analogues expression for $\tilde{K}^{-}_{f(b)}$

\eqnarray
\nonumber
\tilde{K}_{f(b)}^- &= &
\left(
\prod_{p_-}
\mbox{\large e}^{-(\beta - it)\varepsilon_{p_-}}
K_{p_-}(\beta - it)
\prod_{p_+}
\mbox{\large e}^{- it\varepsilon_{p_+}}
K_{p_+}(it)
\right)^{\frac12}
\frac{(-i)}{M}
\sum_{p_-,p_{-}'}
\mbox{\large e}^{- ip_-' - ip_-(L+1)}
\times
\\
& &
\nonumber
\times
\int
\mbox{d} \xi^0\, \mbox{d} \bar{\xi}^0\,
\mbox{d} \bar{\xi}^1\,\mbox{d} \xi^1 \,
\mbox{d} \bar{\xi}^2\, \mbox{d} \xi^2\
\exp
\left\{
\bar{\xi}^0\xi^0 -
\bar{\xi}^1\xi^1 -
\bar{\xi}^2\xi^2
\right\}
\bar{\xi}_{p_-}^1\bar{\xi}_{p_-'}^2
\times
\\
& &
\nonumber
\times
\exp
\left\{
\bar{\xi}^0D_-^{(1)}\xi^1 +
\right.
+\bar{\xi}^0B_-^{(1)}\bar{\xi}^0 -
\xi^1B_-^{(1)}\xi^1 -
\bar{\xi}^1UD_+^{(2)}U^+\xi^2 +
\\
& &
\left.
+\bar{\xi}^1UB_+^{(2)}U^t\bar{\xi}^1 -
\xi^2\bar{U}B_+^{(2)}U^+\xi^2 -
\bar{\xi}^2W_-^{f(b)}\xi^0
\right\}
\label{fK-}
\endeqnarray

The next step is to calculate the integrals in (\ref{fK+},\ref{fK-}).
We will consider just one of them and write another one by analogue.
Let us introduce the generating function for the integral in
$\tilde{K}^{+}_{f(b)}$:

\eqnarray
\nonumber
J_{+}^{f(b)}(\eta_1,\eta_2) &=&
\int
\mbox{d} \xi^0\, \mbox{d} \bar{\xi}^0\,
\mbox{d} \bar{\xi}^1\,\mbox{d} \xi^1\,
\mbox{d} \bar{\xi}^2\,\mbox{d} \xi^2\
\exp
\left\{
\bar{\xi}^0\xi^0 -
\bar{\xi}^1\xi^1 -
\bar{\xi}^2\xi^2
\right\}
\times
\\
& &
\nonumber
\times
\exp
\left\{
\bar{\xi}^0D_+^{(1)}\xi^1 +
\bar{\xi}^0B_+^{(1)}\bar{\xi}^0 -
\xi^1B_+^{(1)}\xi^1 -
\bar{\xi}^1U^+D_-^{(2)}U\xi^2 +
\right.
\\
\nonumber
& &
\left.
+\bar{\xi}^1U^+B_-^{(2)}\bar{U}\bar{\xi}^1 -
\xi^2U^tB_-^{(2)}U\xi^2 -
\bar{\xi}^2W_+^{f(b)}\xi^0 +
\bar{\xi}^1\eta_1 +
\bar{\xi}^2\eta_2
\right\}
\endeqnarray

Integrating in the same way as in the previous section and neglecting
terms with $\eta_1$, $\eta_1$ and $\eta_2$, $\eta_2$ one can obtain

$$
J^{+}_{f(b)}(\eta_1,\eta_2) =
(-1)^{\frac{M(M+1)}{2}}
\mathop{\mbox{Pf}}(\tilde{G}_+^{f(b)})
\sqrt{\det R_-^{-1}}
\exp\left\{ \eta_1 F^{f(b)}_+ \eta_2\right\}
$$
Here matrices $Q$, $S$ and $T$ are the same as in~(\ref{QTS}) and
matrices $\tilde{G}_\pm^{f(b)}$ and $F^{f(b)}_\pm$ are determined by

\eqnarray
\nonumber
\tilde{G}_\pm^{f(b)} &=&
\left(
\array{cc}
S_\pm & -\left( Q_\pm^{f(b)}\right)^t
\\
Q_pm^{f(b)} & T_\pm^{f(b)}
\endarray
\right)
\\
\nonumber
F^{f(b)}_\pm &=&
\left(
2B^{(1)}_\pm +
D^{(1)}_\pm
(W^{f(b)}_\pm)^t
\left(
Q_\pm^{f(b)}
\right)^{-t}
S_\pm
\right)
\left(
Q_\pm^{f(b)} +
T_\pm^{f(b)}
\left(
Q_\pm^{f(b)}
\right)^{-t}
S_\pm
\right)^{-1}
\endeqnarray
Using this generating function we can write expressions for all
$\tilde{K}$ in~(\ref{I++}) applying the same calculations as in the
previous section.  The result is given by the following formula:

\eqnarray
\nonumber
\tilde{K}_\pm^{f(b)} &=&
(-1)^{\frac{M(M+1)}{2}}
\left(
\prod_{p_\pm}
\mbox{\large e}^{- (\beta-it)\varepsilon_{p_\pm}}
K_{p_\pm}(\beta - it)
\prod_{p_\mp}
\mbox{\large e}^{-it\varepsilon_{p_\mp}}
K_{p_\mp}(it)
\right)^{\frac12}
\times
\\
\label{K++}
& &
\times
\mathop{\mbox{Pf}}(\tilde{G}_\pm^{f(b)})
\sqrt{\det R_\mp^{-1}}
\frac{(-i)}{M}
\sum_{p_\pm ,p_\pm'}
\mbox{\large e}^{-ip_\pm' - ip_\pm (L+1)}
\left(F^{f(b)}_{\pm}\right)_{p_\pm p_\pm'}
\endeqnarray
Here and matrices $Q_-$, $S_-$ and $T_-$ are the same as
in~(\ref{QTS-}).  So formula~(\ref{K++}) gives us all the necessary
terms to calculate correlator~(\ref{I++}).

\section{Correlation functions for $\gamma =0$ (XX-model)}

Up to day there are three original works devoted to the calculation
of the correlators for the XX-models at finite temperature.
The
first of them is the paper by Colomo, Izergin, Korepin and Tognetti
\cite{Iz1} where the correlator for the XX-model was calculated in
the thermodynamic limit.
The correlators for finite cyclic chains
were calculated by Suzuura, Tokihiro and Ohta  in \cite{Jap} and by
Ilinski, Kalinin and Kapitonov \cite{IKK} for the qXX-model (which
produces the XX-model for the particular value of the deformation
parameter $q=-1$).
Thus, there are three crucial points for the
verification of the results obtained in this paper. Besides this, it
is possible to compare our result with that of
calculated by the
definition for small chains.
It means to represent the
Hamiltonian and the creation-annihilation operators as matrices
$2^{M}\times2^{M}$ for the $M$-site chain and then to evaluate the
trace of the matrix
$
\mbox{\large e}^{-(\beta-it){\bf H}}P_1
\mbox{\large e}^{-it{\bf H}}P^+_{L+1}
$.
The result of the comparison will be reviewed
in this section.

Let us start with the finite XX-chain correlator. One can extract it
from the formulae (\ref{**}--\ref{R}) putting $\gamma = 0$.  Then
$\theta_p=0$ for all $p$ and $E_p=\varepsilon_p$.
\begin{equation}
\langle P_1(t)\ P^+_{L+1}(0) \rangle =
\frac {1}{2Z} (K^f_+ + K^f_- + K^b_+ - K^b_-)
\label{www}
\end{equation}
\begin{eqnarray}
K_\pm^{f(b)} &=&
(-1)^{M(M-1)/2}
\prod\limits_{p\in A_\pm}
\mbox{\large e}^{-it\varepsilon_p}
\mathop{\mbox{Pf}}
\left(
\begin{array}{cc}
0 & (F^{f(b)}_\pm)^t
\\
-F^{f(b)}_\pm & 0
\end{array}
\right)
\mathop{\mbox{Tr}}
(F^{f(b)^{-1}}_\pm G) =
\nonumber
\\
&=&
\mbox{\large e}^{-it(\epsilon)M}
\det F^{f(b)}_\pm
\mathop{\mbox{Tr}} (F^{f(b)}_\pm G)
\end{eqnarray}
where

\begin{eqnarray}
F^{f(b)}_+ &=&
\left.
\left[
W^{f(b)}_+
D^{(1)}_+ +
U^+ D^{(2)}_- R_- U
\right]
\right|_{\gamma=0}
\nonumber
\\
F^{f(b)}_- &=&
\left.
\left[
W^{f(b)}_-
D^{(1)}_- +
U D^{(2)}_+ R_+ U^+
\right]
\right|_{\gamma=0}
\nonumber
\\
G_{p,p^\prime} &=&
\frac1M\mbox{\large e}^{-ip(L+1) + ip^\prime}
\end{eqnarray}
If we rewrite this formula in the coordinate representation we will
get the formula for the correlator  in the qXX-model obtained in
\cite{IKK} for the case $q=-1$.
\begin{equation}
K_\pm^{f(b)} =
\frac {\mbox{\large e}^{-it\epsilon M}}{2Z}
\det \tilde F^{f(b)}_\pm
(\tilde F^{f(b)}_\pm)^{-1}_{1,L+1}
\label{kkk}
\end{equation}
\begin{eqnarray}
(\tilde F^f_\pm)_{m,m^\prime} &=&
\frac 1M \sum\limits_{r\in A_\mp} \mbox{\large e}^{it\varepsilon_r
+ ir(m-m^\prime)} +
\frac 1M \sum\limits_{p\in A_\pm} \mbox{\large e}^{ip(m-m^\prime)
+ (it-\beta)\varepsilon_p}
\biggl( 1 - 2 \times \left\{
\begin{array}{c}
1 \mbox{\ if\ } m\leq L \\
0 \mbox{\ if\ } m > L
\end{array} \right. \biggr)
\\
(\tilde F^b_\pm)_{m,m^\prime} &=&
\frac 1M \sum\limits_{r\in A_\mp} \mbox{\large e}^{it\varepsilon_r
+ ir(m-m^\prime)} -
\frac 1M \sum\limits_{p\in A_\pm} \mbox{\large e}^{ip(m-m^\prime)
+ (it-\beta)\varepsilon_p}
\biggl( 1 - 2 \times \left\{
\begin{array}{c}
1 \mbox{\ if\ } m\leq L \\
0 \mbox{\ if\ } m > L
\end{array} \right. \biggr)
\label{KK}
\end{eqnarray}

One should note that our result is different from that announced
in \cite{Jap}.  First of all in our opinion there is a misprint in
signs of two last terms in Eqn.(15) of Ref.\cite{Jap}. Moreover, the
numerical calculations for $t = 0$ and particular values of
parameters ($\{\epsilon=1, h=5\},\{\epsilon=1, h=2\}$) reveal
the difference between the results of \cite{Jap} and the calculation
by the definition for both choices of the sign.

Eqs.(\ref{www},\ref{kkk}) coincide with the correlator calculated by
the definition for the number of sites $M=3$. In this case it is
possible to compare the analytical expressions for the correlators
(the calculation was performed using the system of symbolical
calculations REDUCE).

Let us now consider the thermodynamic limit ($M\rightarrow\infty$)
for the XX-model ($\gamma=0$).  At first we will treat the partition
function.  The ``bosonic" contributions $Z^{\pm}_{b}$ are equal and,
hence, disappear.  By the same reason the ``fermionic" terms are
equal and cancel the factor $2$ in the formula (\ref{Z}).  This is
due to the absence of the difference of momentum spectrum $A_{+}$ and
$A_{-}$ (in the thermodynamic limit the boundary conditions are
irrelevant).  So the partition function of the XX-model in the
thermodynamic limit is equal to the partition function of free
fermions on the infinite chain.

In the same way the correlator for the XX-model in the thermodynamic
limit is obtained from Eqn.(\ref{www}) putting
$K^f_+=K^f_-\equiv K$, $K^b_+=K^b_-$.
So in the thermodynamic limit we have
\begin{equation}
        \langle P_0(t)\ P^+_{L}(0) \rangle = \frac{K}{Z}
\label{xxx1}
\end{equation}
where
\begin{equation}
        Z = \prod_p (1+\mbox{\large e}^{-\beta \varepsilon_p} )
\label{xxx2}
\end{equation}
is the fermionic partition function and
\begin{equation}
        K = \mbox{\large e}^{-it\epsilon M} \det(F)
        \frac{1}{M}\sum\limits_{p,p'} \mbox{\large e}^{ipL}
        F^{-1}_{p'p} \ ,
\label{xxx3}
\end{equation}
$$
        F = R(it-\beta) + R(it) - 2 V R(it-\beta)\ ,
$$
$$
        R(\tau) = \mbox{diag}\{\mbox{\large e}^{\tau\varepsilon_p}\}
$$
and the matrix $V$ is given by Eqn.(\ref{V}).

It is easy to show that the matrix $F$ can be represented in the
following form:
$$
        F = \theta^{-1/2} [1-2 \theta^{1/2} V \theta^{1/2}]
        \theta^{-1/2} R(it-\beta)
$$
where $\theta$ is the following diagonal matrix
$$
        \theta = \mbox{diag}\left\{
        \frac{1}{1+\mbox{\large e}^{\beta\varepsilon_p}}
        \right\}\ .
$$
So we obtain
$$
        \det(F) = Z \mbox{\large e}^{it\varepsilon M}
        \det[1-2 \theta^{1/2} V \theta^{1/2}]\ ,
$$
hence
\begin{eqnarray}
        \langle P_0(t)\ P^+_{L}(0) \rangle &=&
        \det[1-2 \theta^{1/2} V \theta^{1/2}]\times\nonumber\\
        &&\nonumber\times
        \frac{1}{M}\sum\limits\limits_{p,p'} \mbox{\large e}^{ipL}
        R_{p'p'}(\beta-it)\theta^{1/2}_{p'p'}
        [1-2 \theta^{1/2} V \theta^{1/2}]^{-1}_{p'p}
        \theta^{1/2}_{pp}\ .
\end{eqnarray}
Let us consider the following unitary transformation of $V$:
$$
        V = U^{-1} v U
$$
where
\begin{equation}
        U = \mbox{diag}\{\mbox{\large e}^{ip(L+1)/2}\}\ ,\quad
        v_{pp'} = \frac{\sin\frac{p'-p}{2}L}{\sin\frac{p'-p}{2}}
\label{xxx4}
\end{equation}
Then the correlator takes the form
\begin{eqnarray}
        \langle P_0(t)\ P^+_{L}(0) \rangle &=&
        \det[1-2 \theta^{1/2} v \theta^{1/2}]
        \frac{1}{M}\sum\limits\limits_{p,p'}
        \mbox{\large e}^{i(p+p')L/2
        +i(p-p')/2} R_{p'p'}(\beta-it)\times
        \nonumber\\
        &&\nonumber\times \theta^{1/2}_{p'p'}
        [1-2 \theta^{1/2} v \theta^{1/2}]^{-1}_{p'p}
        \theta^{1/2}_{pp}\\
        &=&
        \det[1-2 \theta^{1/2} v \theta^{1/2}]
        \frac{1}{M}\sum\limits\limits_{p,p'}
        \theta^{1/2}_{p'p'} [1-2 \theta^{1/2} v
        \theta^{1/2}]^{-1}_{p'p}\theta^{1/2}_{pp}
        \times\nonumber\\
        &&\nonumber\times
        \mbox{\large e}^{i(p+p')L/2}
        \frac{1}{2}
        [\mbox{\large e}^{i(p-p')/2+(\beta-it)\varepsilon_{p'}}
        +\mbox{\large e}^{i(p'-p)/2+(\beta-it)\varepsilon_{p}}]
\end{eqnarray}
And finally we obtain the following expression for the correlator
\begin{eqnarray}
        \langle P_0(t)\ P^+_{L}(0) \rangle &=&
        \det[1-2 \theta^{1/2} v \theta^{1/2}]
        \mathop{\mbox{Tr}} \left\{
        \theta^{1/2} r(\beta-it) \theta^{1/2} [1-2 \theta^{1/2} v
        \theta^{1/2}]^{-1}\right\} \nonumber\\
        &=&
        \left.\frac{\partial}{\partial z}
        \det[1-2 \theta^{1/2} v \theta^{1/2}+z\theta^{1/2}
        r(\beta-it) \theta^{1/2} ]\right|_{z=0}
\label{xxx5}
\end{eqnarray}
where $v$ is defined in (\ref{xxx4}) and
\begin{equation}
        r_{pp'}(\tau) =
        \mbox{\large e}^{i(p+p')L/2}\frac{1}{2}
        [\mbox{\large e}^{i(p-p')/2+\tau\varepsilon_{p'}}
        +\mbox{\large e}^{i(p'-p)/2+\tau\varepsilon_{p}}]
\label{xxx6}
\end{equation}
In the same way expression for correlator
$\langle P^+_0(t)\ P_{L}(0) \rangle$ is obtained
\begin{eqnarray}
        \langle P^+_0(t)\ P_{L}(0) \rangle &=&
        \det[1-2 \theta^{1/2} v \theta^{1/2}]
        \mathop{\mbox{Tr}} \left\{
        \theta^{1/2} r(it) \theta^{1/2} [1-2 \theta^{1/2} v
        \theta^{1/2}]^{-1} \right\} \nonumber\\
        &=&
        \left.\frac{\partial}{\partial z}
        \det[1-2 \theta^{1/2} v \theta^{1/2}+z \theta^{1/2} r(it)
        \theta^{1/2} ] \right|_{z=0}
\label{xxx7}
\end{eqnarray}

The form of expression (\ref{xxx7}) is similar to that in
Ref.\cite{Iz1}.  Let us substitute $t=0$ in Eqn.(\ref{xxx7}):
\begin{eqnarray}
        \langle P^+_0\ P_{L} \rangle &=&
        \det[1-2 \theta^{1/2} v \theta^{1/2}]
        \mathop{\mbox{Tr}} \left\{
        \theta^{1/2} r \theta^{1/2} [1-2 \theta^{1/2} v
        \theta^{1/2}]^{-1}\right\} \nonumber\\
        &=&
        \left.\frac{\partial}{\partial z}
        \det[1-2 \theta^{1/2} v \theta^{1/2}+z \theta^{1/2} r
        \theta^{1/2} ]\right|_{z=0}
\label{xxx9}
\end{eqnarray}
where
\begin{equation}
        r_{pp'} \equiv r_{pp'}(0) =
        \mbox{\large e}^{i(p+p')L/2}\cos\frac{(p-p')}{2}
\label{xxx10}
\end{equation}

Now we can see that expression (\ref{xxx9}) is different from
Eqn.(39) of Ref.\cite{Iz1} in two aspects. At first, the matrix $r$
contains the additional factor $\cos\frac{(p-p')}{2}$. Secondary, in
the matrix $v$ we should replace $L$ by $L-1$. We will show that
these two expressions for the correlator
$\langle P^+_0\ P_{L} \rangle$ are the same.
Indeed, we can write down from the definition
\begin{eqnarray}
        \langle P^+_0\ P_{L} \rangle &=&
        \frac{1}{Z}\mathop{\mbox{Tr}} \left[
        \mbox{\large e}^{-\beta {\bf H}} P^+_0 P_L \right] \nonumber\\
        &=& \frac{1}{Z}\int\mbox{d}\xi_0\,\mbox{d}\bar{\xi}_0\,
        \mbox{d}\bar{\xi}_1\, \mbox{d}\xi_1\
	\exp[\bar{\xi}_0 \xi_0 - \bar{\xi}_1 \xi_1]\times
        \nonumber\\
        &&\times
        \langle\bar{\xi}_0|\mbox{\large e}^{-\beta {\bf H}}|\xi_1\rangle
	\langle\bar{\xi}_1|\mathop{\mbox{N}}c^+_0
        \exp\{-2\sum\limits\limits_{k=1}^{L-1}c_k^+c_k\} c_L |\xi_0
        \rangle\ .
\label{xxx11}
\end{eqnarray}
Using the method described in Section~3 we obtain the following result
\begin{eqnarray}
        \langle P^+_0\ P_{L} \rangle &=&
        \det[1-2 \theta^{1/2} \tilde v \theta^{1/2}]
        \mathop{\mbox{Tr}} \left\{
        \theta^{1/2} \tilde r \theta^{1/2} [1-2 \theta^{1/2} \tilde v
        \theta^{1/2}]^{-1} \right\} \nonumber\\
        &=&
        \left.\frac{\partial}{\partial z}
        \det[1-2 \theta^{1/2} \tilde v \theta^{1/2}+z \theta^{1/2}
        \tilde r \theta^{1/2}]\right|_{z=0}
\label{xxx12}
\end{eqnarray}
where
$$
        \tilde{r}_{pp'} = \mbox{\large e}^{i(p+p')L/2}
$$
$$
        \tilde{v}_{pp'} =
        \frac{\sin\frac12(p'-p)(L-1)}{\sin\frac12(p'-p)}
$$
As $c^+_0 c_L=c^+_0 \exp[i\pi c^+_0c_0] c_L$ we can rewrite
(\ref{xxx11}) in the following form
\begin{eqnarray}
        \langle P^+_0\ P_{L} \rangle &=&
        \frac{1}{Z}\int\mbox{d}\xi_0\,\mbox{d}\bar{\xi}_0\,
        \mbox{d}\bar{\xi}_1\, \mbox{d}\xi_1\
	\exp[\bar{\xi}_0 \xi_0 - \bar{\xi}_1 \xi_1]\times
        \nonumber\\
        &&\times
        \langle\bar{\xi}_0|\mbox{\large e}^{-\beta {\bf H}}|\xi_1\rangle
	\langle\bar{\xi}_1|\mathop{\mbox{N}}c^+_0
        \exp\{-2\sum\limits\limits_{k=0}^{L-1}c_k^+c_k\} c_L |\xi_0
        \rangle\ .
\label{xxx13}
\end{eqnarray}
Repeating the above calculation we can come to expression
(\ref{xxx9}).  It follows from the previous consideration that
expressions (\ref{xxx9}) and (\ref{xxx12}) are in fact the same.

So it is not now surprising that expression (\ref{xxx7}) for correlator
$\langle P^+_0(t)\ P_{L}(0) \rangle$ is different from that in
Ref.\cite{Iz1} because they are derived with different methods and as
we have seen above it is difficult to transform one form to another.
However, due to the verification for finite $M$ and for $t=0,
M\rightarrow\infty$ we can be sure in the correctness of our formulae
in $M\rightarrow\infty$ limit and $t\neq0$.  Finally, our result is
obviously simpler. Indeed, expressions (\ref{xxx5},\ref{xxx7}) for
the correlators contain the time dependence only in the matrix $r$
and trivially generalize the $t=0$ case.

\section{Conclusion}

Now we can summarize the results of the previous sections.  Formula
(\ref{*1}) together with  formulae~(\ref{fb}, \ref{**}, \ref{I++},
\ref{K++}) for the correlation functions give the final expression
for the linear response of the molecular aggregate.  In our opinion,
the result has the following advantages:
\begin{enumerate}
\item It can be used for the numerical calculations to investigate
the finite size effects and the closure effects for cyclic
aggregates.  As we could note during computer calculations for the
XX-model, to deal with our expressions is significantly
easier than with the corresponding formulae of Ref.\cite{Jap}
\item We can say the same about the resulting representations for the
correlators in the thermodynamic limit.  We cannot check our
expressions directly in this case but the positive results for  small
finite chains assures us.
\end{enumerate}
Besides, we want to add that our Fredholm determinant representation
of  two-point correlation functions of the XY-model also allows to
write differential equations and to calculate their asymptotic by the
standard way \cite{KIB}.

In conclusion, in the paper we  considered the linear response
function of a gas of cyclic aggregates in the framework of Frenkel
excitons.  To do this we reminded that the problem is effectively
equivalent to the calculation of the correlation functions for the
cyclic 1-D XY-model.  We  obtained the exact formulae for two-point
correlation functions at finite temperature for the model and
compared them with previous results.  In particular, we showed that
for the XX-model our result for the correlation function
$<P_{n}(t)P_{m}^{+}>$ coincides with the straightforward calculation
(using the definition of the correlation function) for the simplest
nontrivial case ($M=3$) and is reduced in general to the result of
paper \cite{IKK} for an arbitrary number of sites.  We also found
that the result of  Ref.\cite{Jap} for the correlation function is
incorrect.  The thermodynamical limit of our formulae was discussed.

We wish to thank V.M.Agranovich and V.S.Kapitonov for the valuable
discussions.  This work was supported by the Russian Fund of
Fundamental Investigations, N 95-01-00548 and partially (K.I.) by the
UK EPSRC under Grant number GR/J35221, Euler stipend of German
Mathematical Society and grant INTAS-939, (V.M.) by President stipend,
(A.S.) ICFPM Fellowship, ISF grant No~A602-F.

\newpage

\appendix
\section{Appendix}

In this appendix we will calculate integrals (\ref{I-}) using the
following formula:

\equation
\label{Int}
\int \mbox{d} \phi \
\exp\left(
\frac12\phi K\phi +
\phi a
\right) =
\mathop{\mbox{Pf}}(K)
\exp\left(
\frac12
aK^{-1}a
\right)
\endequation
where $\phi$ and $a$ are Grassmann vectors and
$\mathop{\mbox{Pf}}(K)$ is a Pfaffian of matrix $K$.

Let us consider $I_{f(b)}^{+}$ and integrate over
$\xi^{0},\bar{\xi}^{0}$ first, using formula (\ref{Int}), where we
will put

$$
\phi =
\left(
\array{c}
\xi^0 \\ \bar{\xi}^0
\endarray
\right)
$$
Then the result of the integration is given by:

$$
\exp\left\{
\bar{\xi}^2W_+^{f(b)}B_+^{(1)}
\left( W_+^{f(b)} \right)^t    \bar{\xi}^2 -
\bar{\xi}^2W_+^{f(b)}D_+^{(1)} \bar{\xi}^1
\right\}
$$
Now we will integrate
the remaining expression over $\xi^{2},\bar{\xi}^{1}$
to keep aside the terms
with $\eta, \bar{\eta}$.
The result takes the form:

$$
\sqrt{\det R_-^{-1}}
\exp
\left\{
\bar{\xi}^2U^+ B_-^{(2)} R_- \bar{U} \bar{\xi}^2 -
\xi^1 U^t B_-^{(2)} R_- U \xi^1 -
\bar{\xi}^2 U^+ (D_-^{(2)})^{-1} R_- U \xi^1
\right\}
$$
where for a convenience we have introduced the diagonal matrix
$R$~(\ref{R}).  The third integration over $\xi^{1},\bar{\xi}^{2}$
gives us the final result.  We will neglect the terms in the exponent
containing $\bar{\eta},\bar{\eta}$  or $\eta ,\eta$ because they will
vanish if we calculate the integrals in $K^{+}_{f(b)}$ using
generating function $I^{+}_{f(b)}(\bar{\eta},\eta)$  (by taking
derivatives on $\eta$ and $\bar{\eta}$).  So we obtain the following
expression:

$$
I^{+}_{f(b)}(\bar{\eta},\eta) =
(-1)^{\frac{M(M-1)}{2}}
\mathop{\mbox{Pf}}(G_+^{f(b)})
\sqrt{\det R_-^{-1}}
\exp
\left\{
\bar{\eta}
\left(
Q_+^{f(b)} +
T_+^{f(b)}
\left(
Q_+^{f(b)}
\right)^{-t}
S_+
\right)^{-1}
\eta
\right\}
$$
where

$$
G_+^{f(b)} =
\left(
\array{cc}
S_+ & \left( Q_+^{f(b)}\right)^t
\\
-Q_+^{f(b)} & T_+^{f(b)}
\endarray
\right)
$$
and for the sake of brevity we have introduced the new matrices
$Q_+^{f(b)}, T_+^{f(b)}$ and $S_+$ which are determined
by~(\ref{QTS}).

In the same manner we can present a result for the generating
function $I^{-}_{f(b)}$:

$$
I^-_{f(b)}(\bar{\eta},\eta) =
(-1)^{\frac{M(M-1)}{2}}
\mathop{\mbox{Pf}}(G_-^{f(b)})
\sqrt{\det R_+^{-1}}
\exp
\left\{
\bar{\eta}
\left(
Q_-^{f(b)} +
T_-^{f(b)}
\left(
Q_-^{f(b)}
\right)^{-t}
S_-
\right)^{-1}
\eta
\right\}
$$
Here $G^{f(b)}_{-}$ is defined by changing all indices ``$+$'' at
operators by indices ``$-$'' in $G^{f(b)}_{+}$ and $Q_-^{f(b)},
T_-^{f(b)}, S_-$ are determined in~(\ref{QTS-}).

To obtain an expression for the integrals in $K^{\pm}_{f(b)}$ one
should consider

$$
\frac{\partial}{\partial\bar{\eta}_{p_+'}}
\frac{\partial}{\partial\eta_{p_+}}
I^\pm_{f(b)}(\bar{\eta}, \eta)
$$
and then the final result for $K^{\pm}_{f(b)}$ can be written as:

\eqnarray
& &
\nonumber
K^\pm_{f(b)} =
(-1)^{\frac{M(M-1)}{2}}
\mathop{\mbox{Pf}}(G_\pm^{f(b)})
\sqrt{\det R_{\mp}^{-1}}
\times
\\
\nonumber
& &
\left(
\prod_{p_\pm}
\mbox{\large e}^{- (\beta-it)\varepsilon_{p_\pm}}
K_{p_\pm}(\beta - it)
\prod_{p_\mp}
\mbox{\large e}^{-it\varepsilon_{p_\mp}}
K_{p_\mp}(it)
\right)^{\frac12}
\times
\\
& &
\nonumber
\frac1M
\sum_{p_\pm ,p_\pm'}
\mbox{\large e}^{ip_\pm' - ip_\pm (L+1)}
        \left(
                Q_\pm^{f(b)} +
                T_\pm^{f(b)}
                \left(
                        Q_\pm^{f(b)}
                \right)^{-t}
                S_\pm
        \right)_{p_\pm'p_\pm}^{-1}
\endeqnarray


\newpage

\begin{thebibliography}{99}
\bibitem{Knoester}
F.S.Spano, J.Knoester, in {\it Advances in magnetic and optical
resonance}
{\bf 18}, 117 (1994);
\bibitem{LMS}
E.Lieb, T.Schults, D.Mattis, {\it Ann.Phys.}(N.Y.)
{\bf 16}, 407, (1961);
\bibitem{Note1} Exact treatment of nonlinear response is in
our forthcoming paper.
\bibitem{IKK} K.N.Ilinski, G.V.Kalinin, V.S.Kapitonov,
preprint hep-th 9404038;
\bibitem{N}
Th.Niemeijer, {\it Physica} {\bf 36}, 377 (1967);
\bibitem{BM}
E.Barouch, B.McCoy {\it Phys.Rev} {\bf A 3}, 786 (1971);
\bibitem{Iz1}
F.Colomo, A.G.Izergin, V.E.Korepin, V.Tognetti, {\it Phys.Lett.}
{\bf A 169}, 243 (1992);
\bibitem{Iz2}
A.R.Its,A.G.Izergin, V.E.Korepin, N.A.Slavnov, {\it Phys.Rev.Lett.}
{\bf 70}, 1704 (1993); {\it Algebra and Analysis} {\bf 6}, N2 (1994)
(in Russian);
\bibitem{Jap}
H.Suzuura, T.Tokihiro, Y.Ohta,
{\it Phys.Rev.} {\bf B 49}, N 6, 4344 (1994);
\bibitem{A0}
V.M.Agranovich, Zh.ETF {\bf 37}, 430 (1959);
\bibitem{KI}
V.S.Kapitonov, K.N.Ilinski, {\it Zapiski Seminarov POMI}, v.224,
72, (1994) (in Russian);
\bibitem{KIB} V.E.Korepin, A.G.Izergin, and N.M.Bogoliubov,
{\it Quantum inverse scattering method, correlation functions and
algebraic Bethe ansatz}, Cambridge Univ. Press, Cambridge, 1992;
\end{thebibliography}
\end{document}